\documentclass[journal=jacsat]{achemso}

\usepackage{amsfonts}
\usepackage{graphicx}
\usepackage{subfigure}

\usepackage[version=3]{mhchem} 
\usepackage[T1]{fontenc}       

\author{Yaozhuang Nie}
\affiliation{School of Physics and Electronics,
Central South University, Changsha,  410083 China}
\email{yznie@csu.edu.cn}

\author{Mavlanjan Rahman}
 \affiliation{School of Physics and Electronics,
Central South University, Changsha,  410083 China}

\author{Daowei Wang}
 \affiliation{School of Physics and Electronics,
Central South University, Changsha,  410083 China}

\author{Can Wang}
 \affiliation{School of Physics and Electronics,
Central South University, Changsha,  410083 China}

\author{Guanghua Guo}
\affiliation{School of Physics and Electronics,
Central South University, Changsha,  410083 China}
\email{guogh@mail.csu.edu.cn}


\title{Strain induced topological phase transitions in monolayer honeycomb structures of group-V binary compounds}

\begin{document}
\begin{tocentry}

\includegraphics[width=7.5cm]{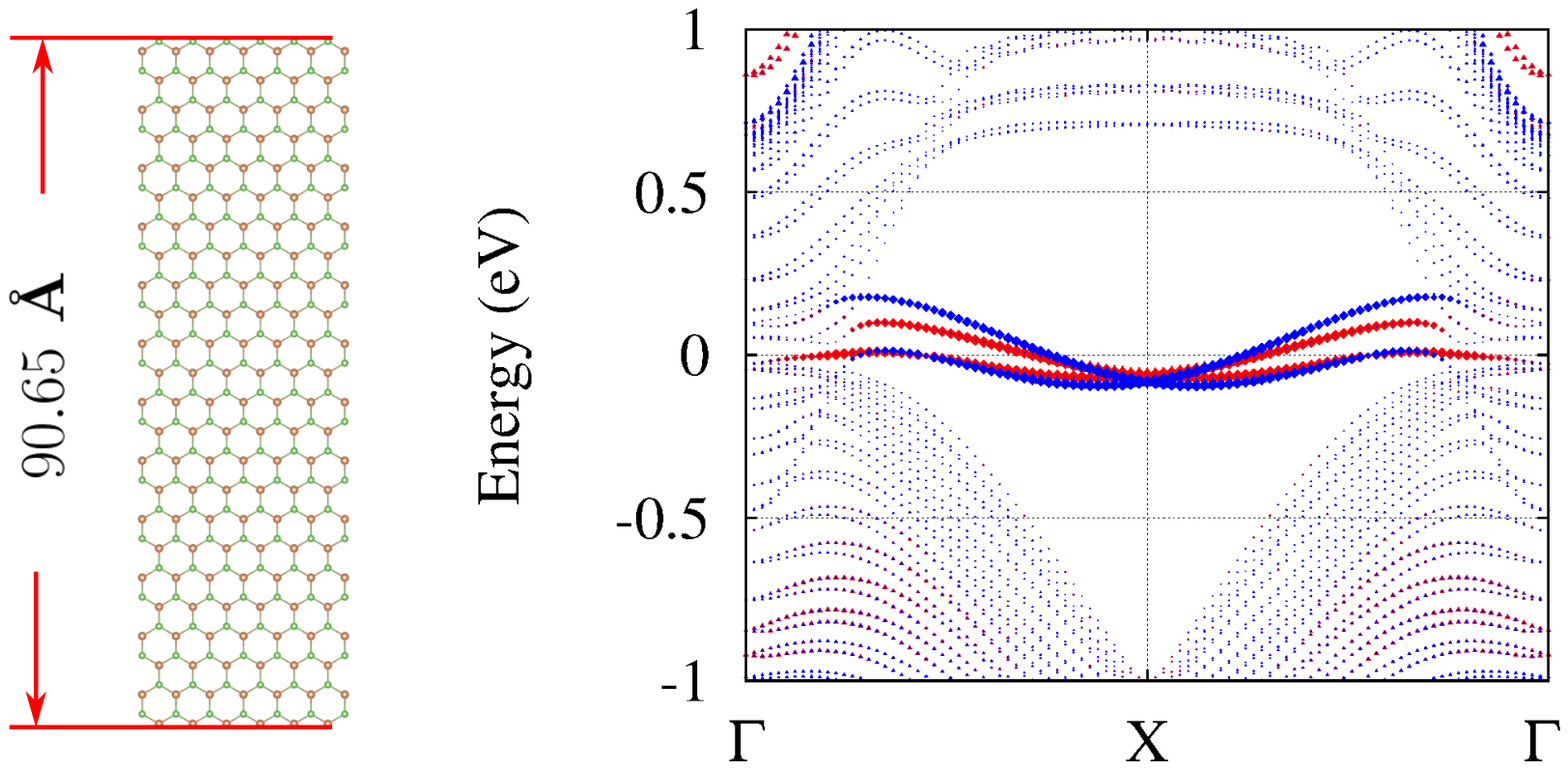}

\end{tocentry}

\begin{abstract}

We present first-principles calculations of electronic structures of a class of two-dimensional (2D) honeycomb structures of group-V binary compounds. Our results show these new 2D materials are stable semiconductors with direct or indirect band gaps. The band gap can be tuned by applying lattice strain. During their stretchable regime, they all exhibit metal-indirect gap semiconductor-direct gap semiconductor-topological insulator (TI) transitions with increasing strain from negative (compressive) to positive (tensile) values. The topological phase transition results from the band inversion at $\Gamma$ point due to lattice strain and is irrelevant to spin-orbit coupling (SOC).

\end{abstract}

Two-dimensional (2D) topological insulators (TIs), also known as quantum spin Hall (QSH) insulators, have attracted much attention recent years due to their rich physics and promising applications in spintronics and quantum computations\cite{hasan_topological_2010,moore_topological_2007,bernevig_quantum_2006,qi_topological_2011}.  The QSH effect, which was first proposed in graphene\cite{kane_z2_2005}, describes the existence of the edge states on the sides of a 2D TI system. These edge states are supposed to consist of two counter-propagating oppositely spin polarized edge channels in the band gap of the material. However, the spin-orbit coupling  (SOC) in graphene is too weak to open a gap large enough to support accessible QSH effect experimentally. Unlike graphene, its elemental analogues - silicene and germanene, show energy gaps because they have larger SOC due to their buckled honeycomb structures\cite{cahangirov_two-_2009}. Therefore, QSH effect can be observed in an experimentally accessible temperature regime in both systems\cite{Liu}. Other monolayer honeycomb structures of group-IV elements as well as III-V binary compounds have been systematically investigated based on first-principles calculations\citep{sahin_monolayer_2009}. Theoretical studies show that many of these materials are 2D TIs at the equilibrium structure, including stanene\citep{liu_low-energy_2011}, InBi, GaBi, and TlBi\cite{chuang_prediction_2014, huang_nontrivial_2014}. Most recently, there has been rising interest in layered compounds of group-V elements\citep{guan_phase_2014,zhu_semiconducting_2014,peng_strain-engineered_2014,fei_strain-engineering_2014, lee_two-dimensional_2014}. Elemental phosphorus have a large number of allotypes. Among them, blue phosphorous has the same layered structure as bulk arsenic, antimony, and bismuth. Their corresponding monolayer materials have the same buckled honeycomb structure as silicene and germanene\cite{guan_phase_2014,kamal_arsenene:_2015}. Monolayer bismuth is reported a 2D TI, while other monolayer buckled honeycomb structures of group-V elements are conventional semiconductors\cite{liu_stable_2011,yao_evolution_2013,wada_localized_2011}.  
Furthermore, recent studies indicate that band topology in these 2D materials could be altered by chemical adsorptions\cite{song_quantum_2014, ma_robust_2015, jin_quantum_2015}, external electric field\cite{ezawa_topological_2012, wang_topological_2013}, and lattice strain\cite{wang_topological_2013,chuang_tunable_2013, huang_strain_2014}. Under lattice strain, monolayer honeycomb structure of P, As, and Sb all exhibit trivial semiconductor to TI transition\cite{huang_strain_2014}. However, the mechanism of the topological phase transition under lattice strain has not been revealed. We also note that there is no report of monolayer honeycomb structures of group-V binary compounds, although bulk AsSb has the same layered structure as bulk arsenic, antimony, and bismuth\cite{kou_structural_2015}. 

In this paper, we study monolayer honeycomb structures of group-V binary compounds (except nitrogen) by first-principles calculations. We investigate the stability of these new 2D materials by studying phonon dispersion and molecular dynamical simulations, and calculate their band structures under lattice strains. We find all of them are stable semiconductors with indirect or direct energy gaps. Under strain, they all exhibit metal-indirect gap semiconductor-direct gap semiconductor-topological insulator (TI) transitions. It is noteworthy that the mechanism of topological transition is due to the different response of energies of the valence band and the conduct band at $\Gamma$ point under lattice strain, and is irrelevant to SOC, which is previously supposed to be indispensable for realizing TIs.

Our first-principles calculations are carried out in the framework of
density-functional theory (DFT)\citep{Kohn:65} within the Perdew-Burke-Ernzerhof generalized gradient approximation (GGA)\cite{PBE} implemented in the ABINIT codes\cite{Gonze}. Norm-conserving HGH pseudopotentials\cite{HGH} and the plane-wave cutoff energy of 30 Hartree are used. For monolayer binary compounds, a supercell with a vacuum space of 20 \AA\ along the z-direction is employed with a $12\times12\times1$ k-point mesh.  Both the lattice parameters and the positions of all atoms are relaxed until the force is less than 1 meV/\AA. The phonon frequencies are computed using density-functional perturbation theory (DFPT)\cite{Baroni} with a $6\times6\times1$ q-point mesh. We perform canonical molecular dynamics (MD) simulation at 300K with a supercell of 72 atoms. The length of time-step of 3 fs and simulations with 500 steps are executed.

\begin{figure}[htbp]
\subfigure[]{\label{structure:a}
\includegraphics[width=5cm]{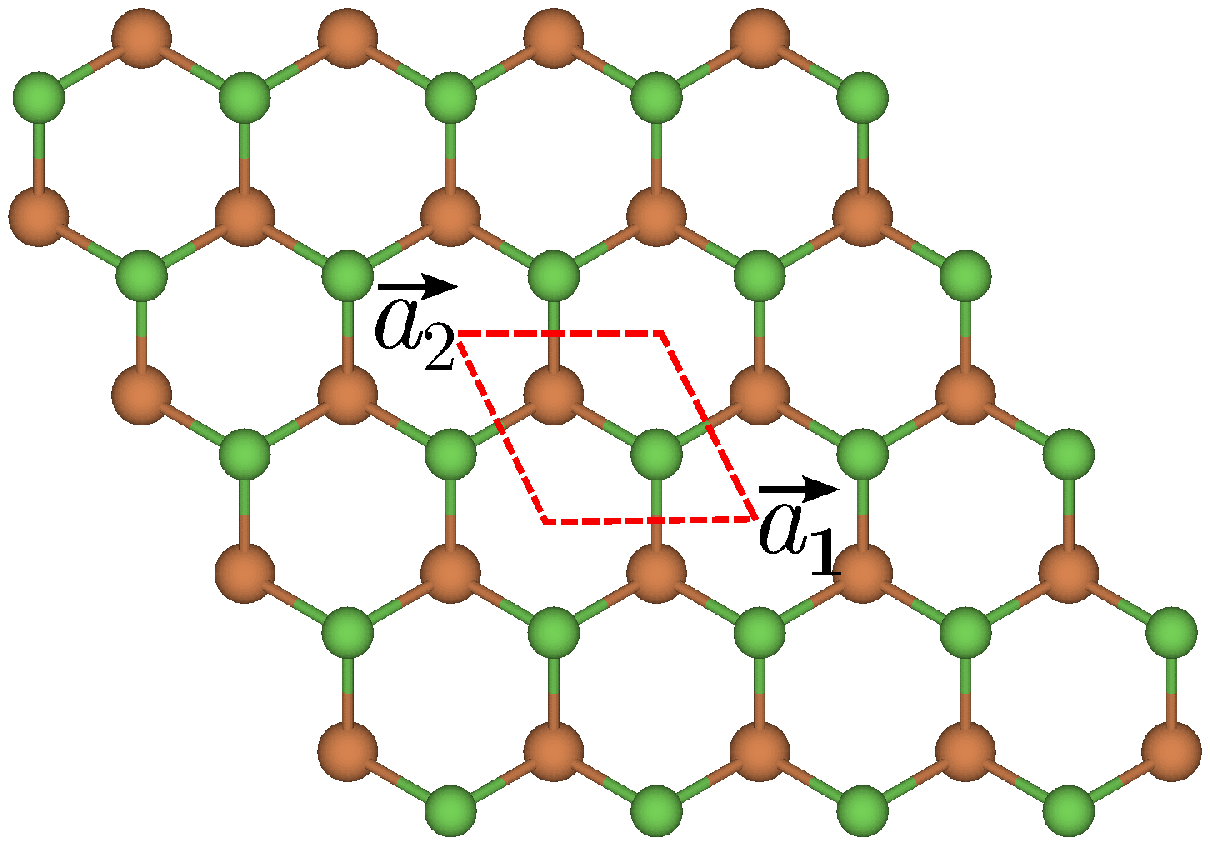}}
\subfigure[]{\label{structure:b}
\includegraphics[width=5cm]{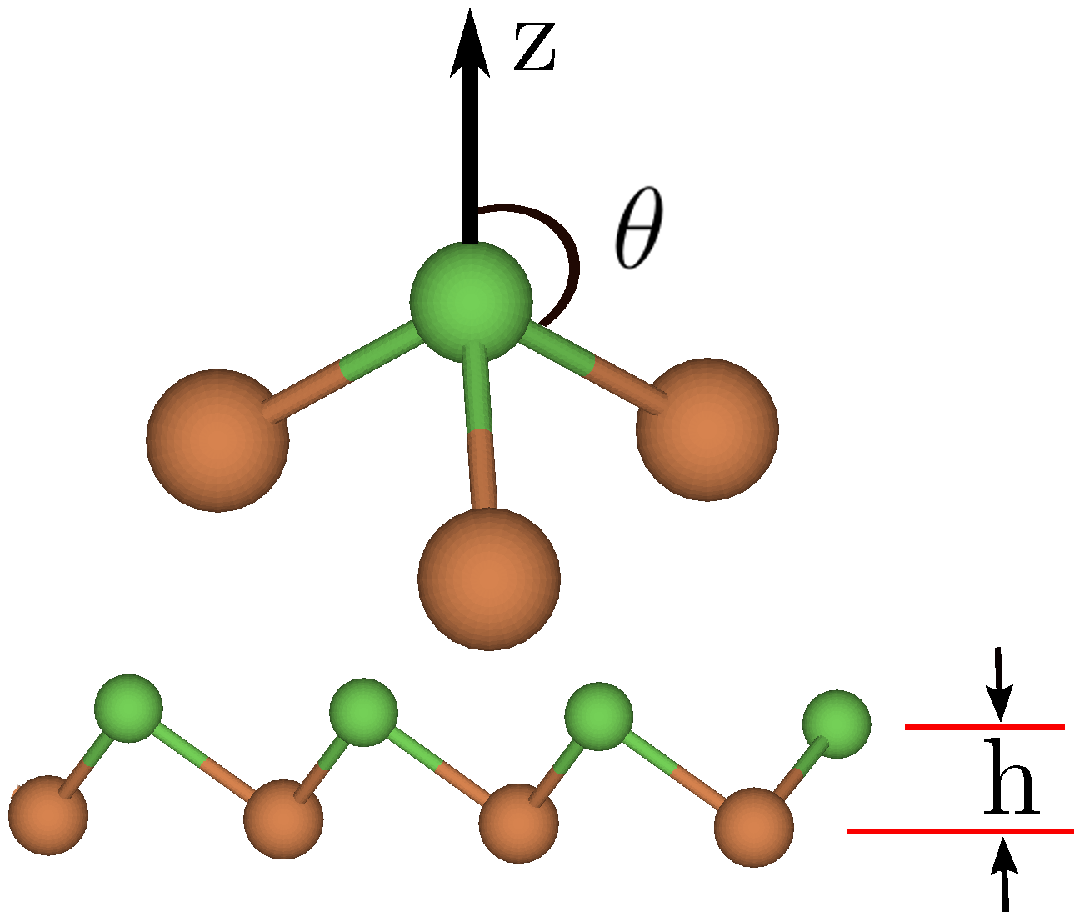}}
\subfigure[]{\label{structure:c}
\includegraphics[width=5cm]{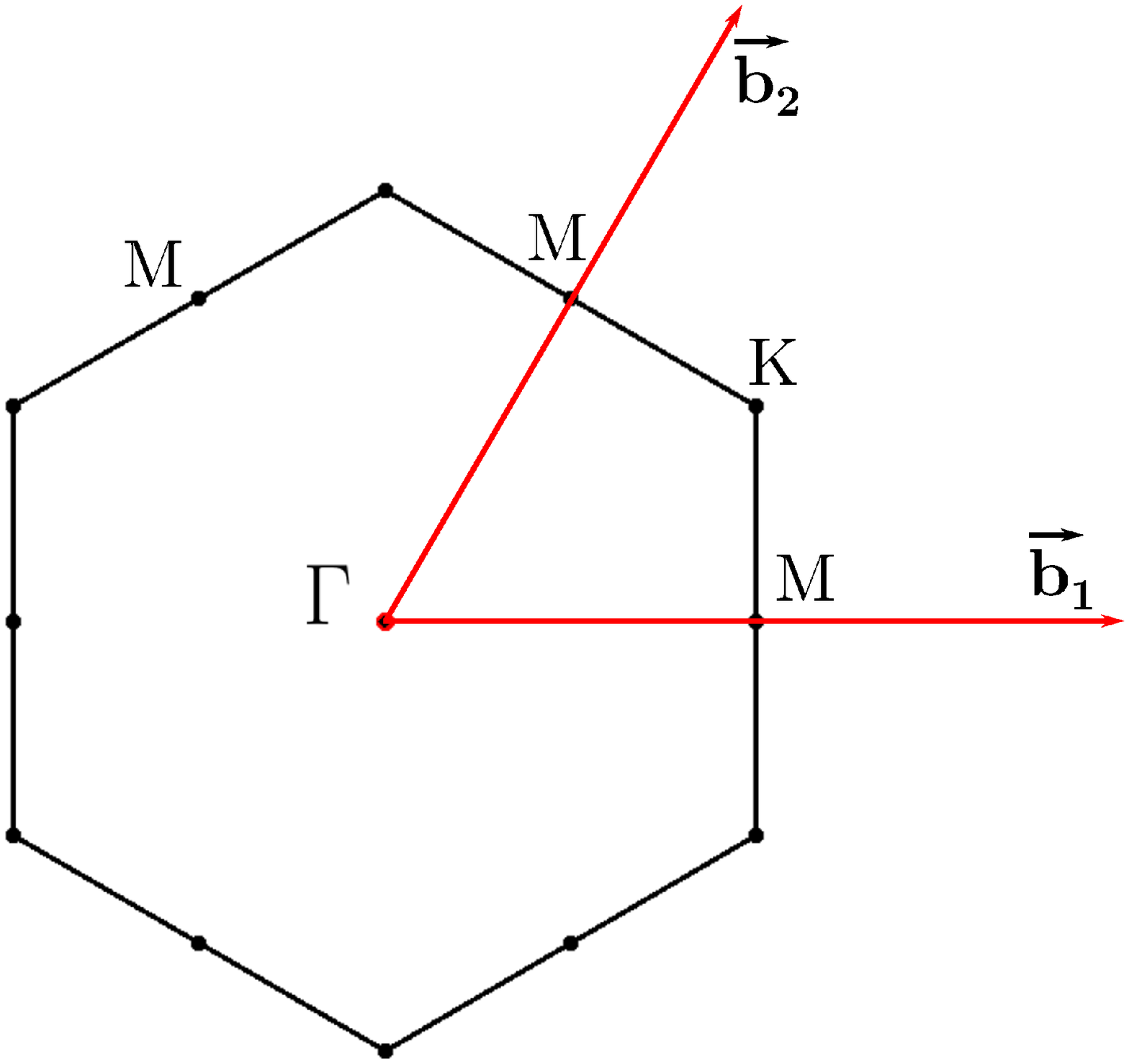}}
\caption{\label{structure} The lattice geometry of buckled honeycomb structure. a: Top view of buckled honeycomb structure. Note that A sublattice and B sublattice (denoted by green and brown color, respectively) are not coplanar. $\vec{a}_{1}=a(1,0)$, $\vec{a}_{2}=a(-\frac{1}{2},\frac{\sqrt{3}}{2})$. b: Side view of buckled honeycomb structure. Definition of the angle $\theta$ and buckling parameter h. c: Brillouin zone and specific symmetry points. $\vec{b}_{1}=\frac{2\pi}{a}(1,\frac{\sqrt{3}}{3})$, $\vec{b}_{2}=\frac{2\pi}{a}(0,\frac{2\sqrt{3}}{3})$}
\end{figure}

Bulk AsSb, or arsenic antimonide, has the same layered crystal structure as arsenic, antimony, and bismuth (space group R$\overline{3}$m, No.166), with the intermediate values of lattice parameters of arsenic and antimony, $a=4.025$ \AA, $c=10.84$ \AA. Like monolayer As and Sb, we consider monolayer AsSb consisting of a layer of As and a layer of Sb. This new material has the same structure as silicene and germinene, i.e. the buckled honeycomb structure, as shown in Figure \ref{structure}. We perform geometry optimization for such a monolayer AsSb, obtain the lattice constants $a=3.867$ \AA, and calculate the binding energy. 
Motivated by this, we consider all possible group-V binary compounds (except nitrogen) with  the same structure. The optimized lattice constants, buckled parameters, and binding energies are calculated, and the corresponding parameters of monolayer P, As ,Sb, and Bi are also given for the sake of comparison (see Supporting Information for more detail). 

\begin{figure}[htbp]
\subfigure[{  } PAs ]{\label{phonon_PAs}
\includegraphics[width=5cm]{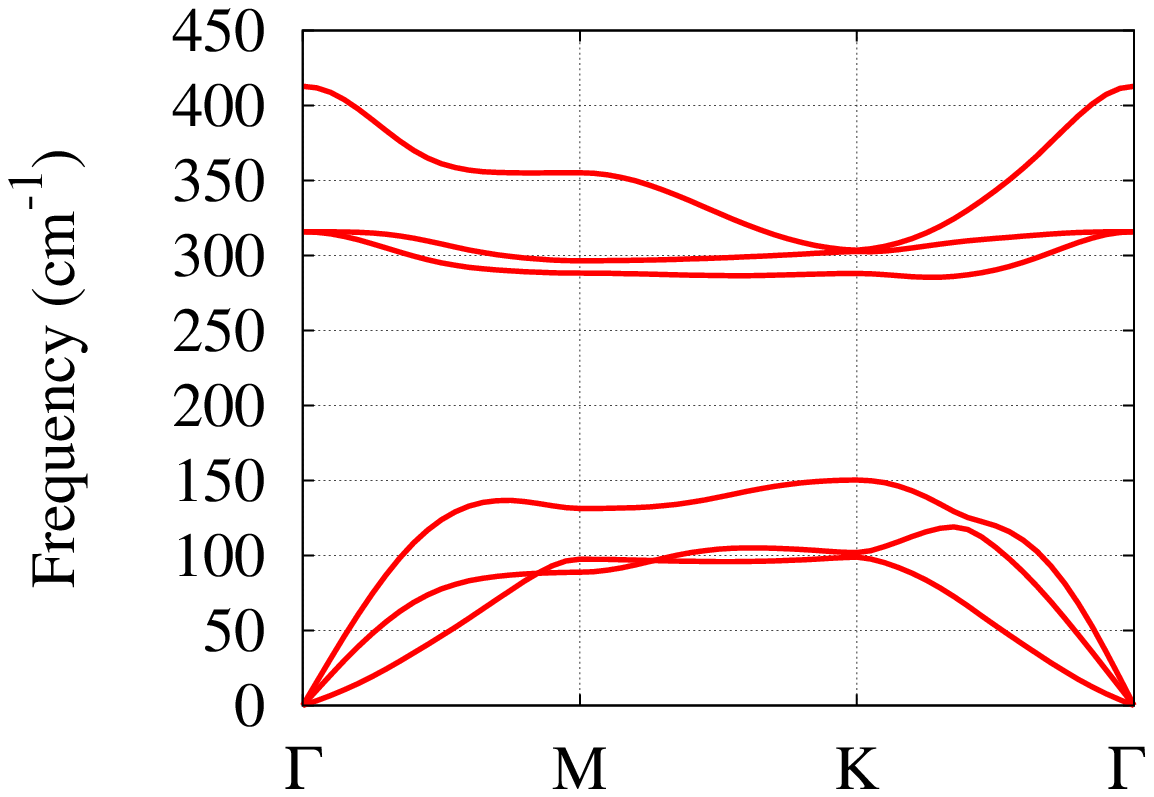}}
\subfigure[ {  } PSb ]{\label{phonon_PSb}
\includegraphics[width=5cm]{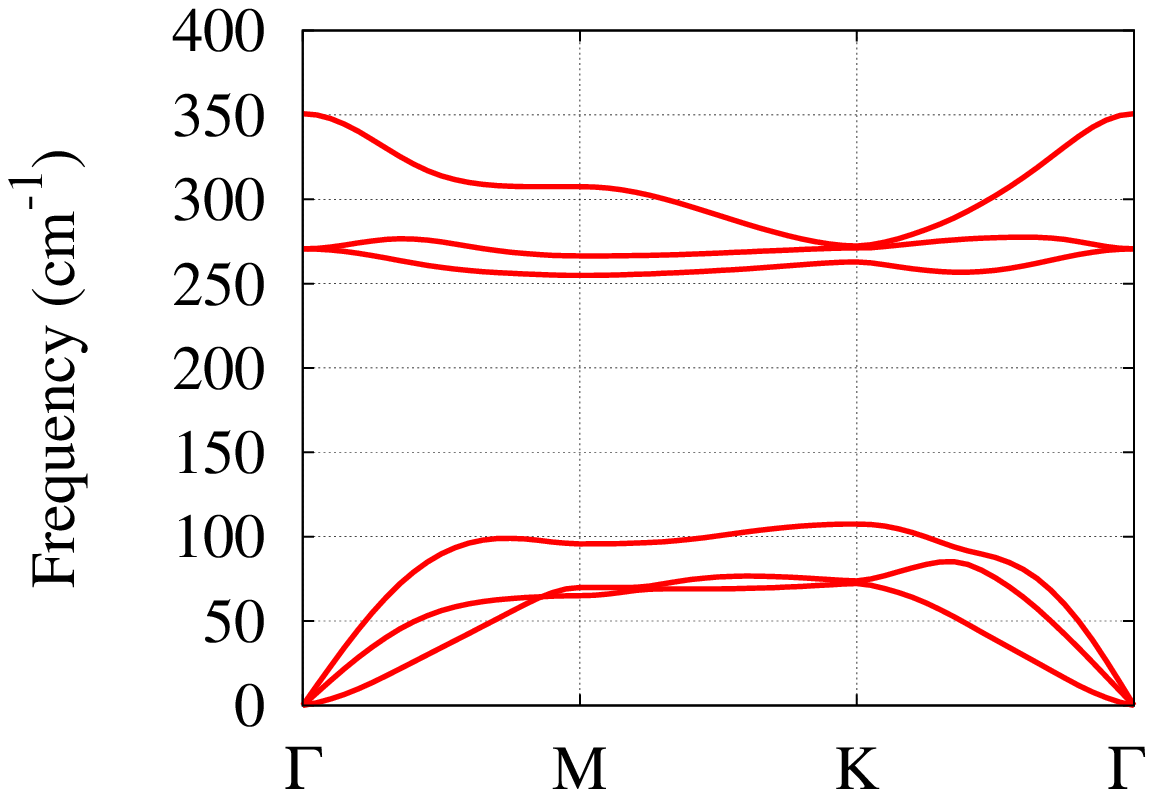}}
\subfigure[ {  } PBi ]{\label{phonon_PBi}
\includegraphics[width=5cm]{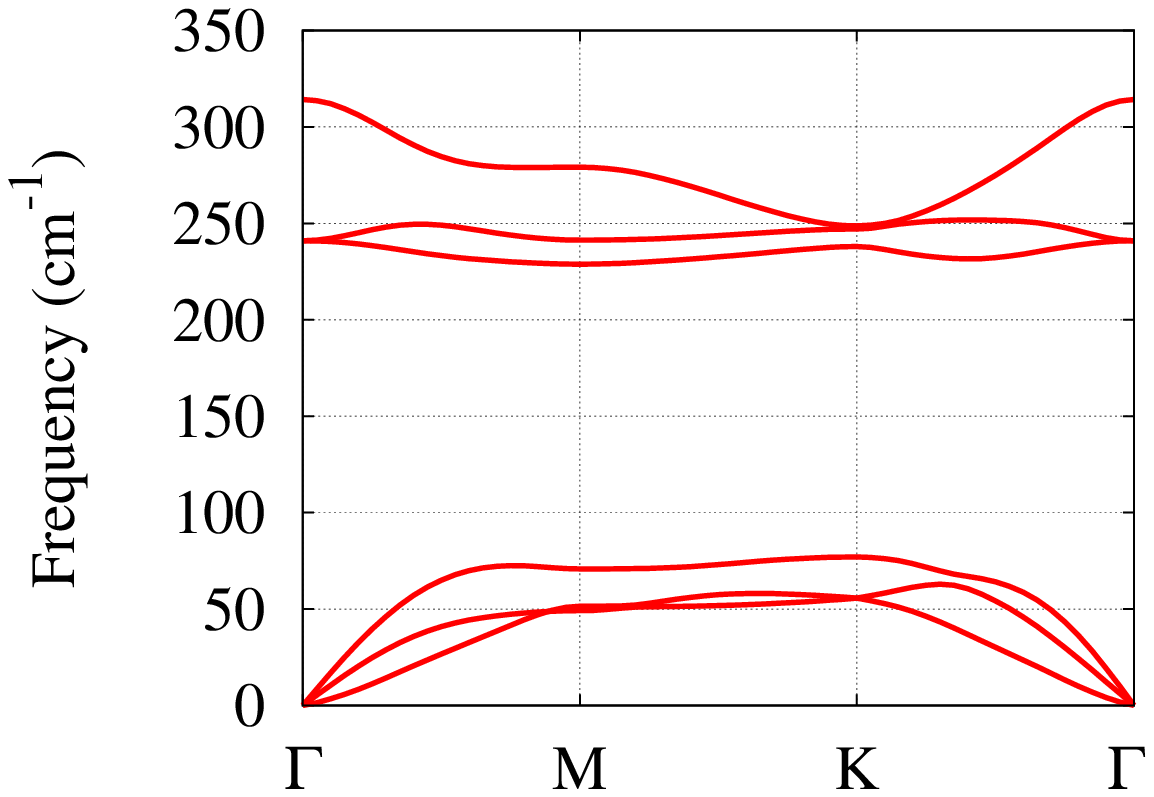}}
\subfigure[{  } AsSb ]{\label{phonon_AsSb}
\includegraphics[width=5cm]{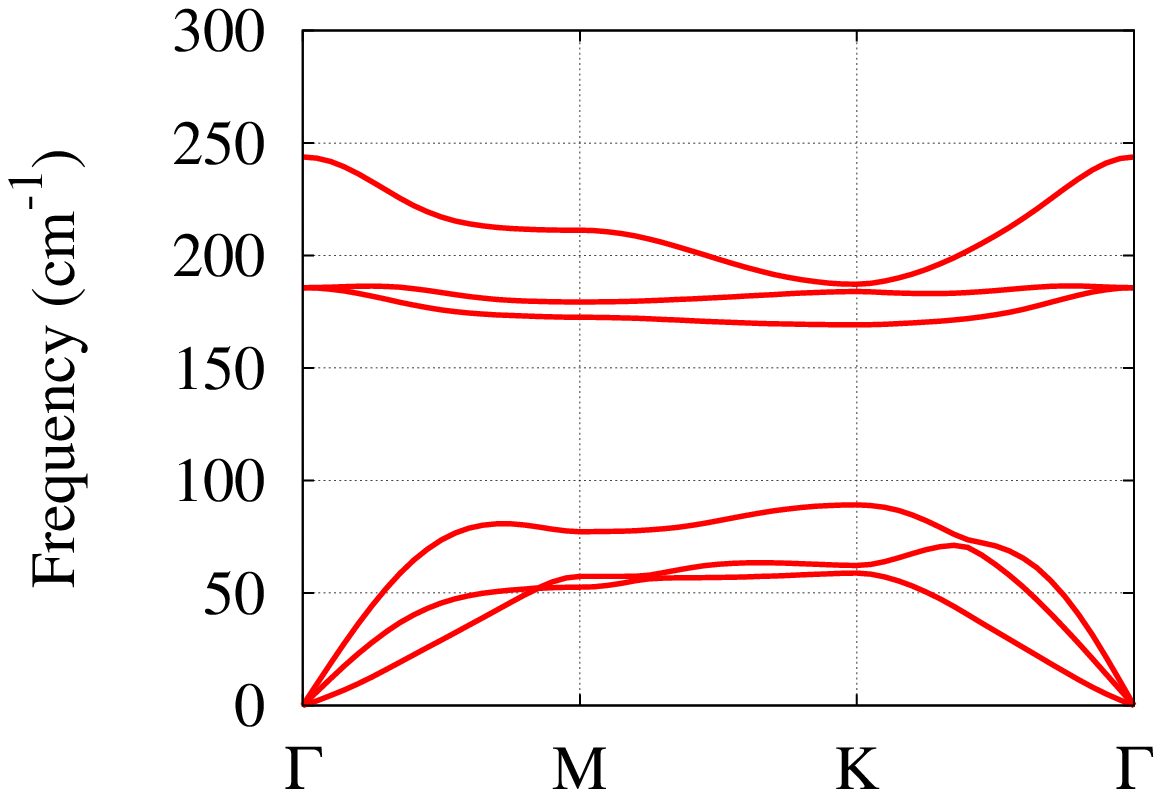}}
\subfigure[ {  } AsBi ]{\label{phonon_AsBi}
\includegraphics[width=5cm]{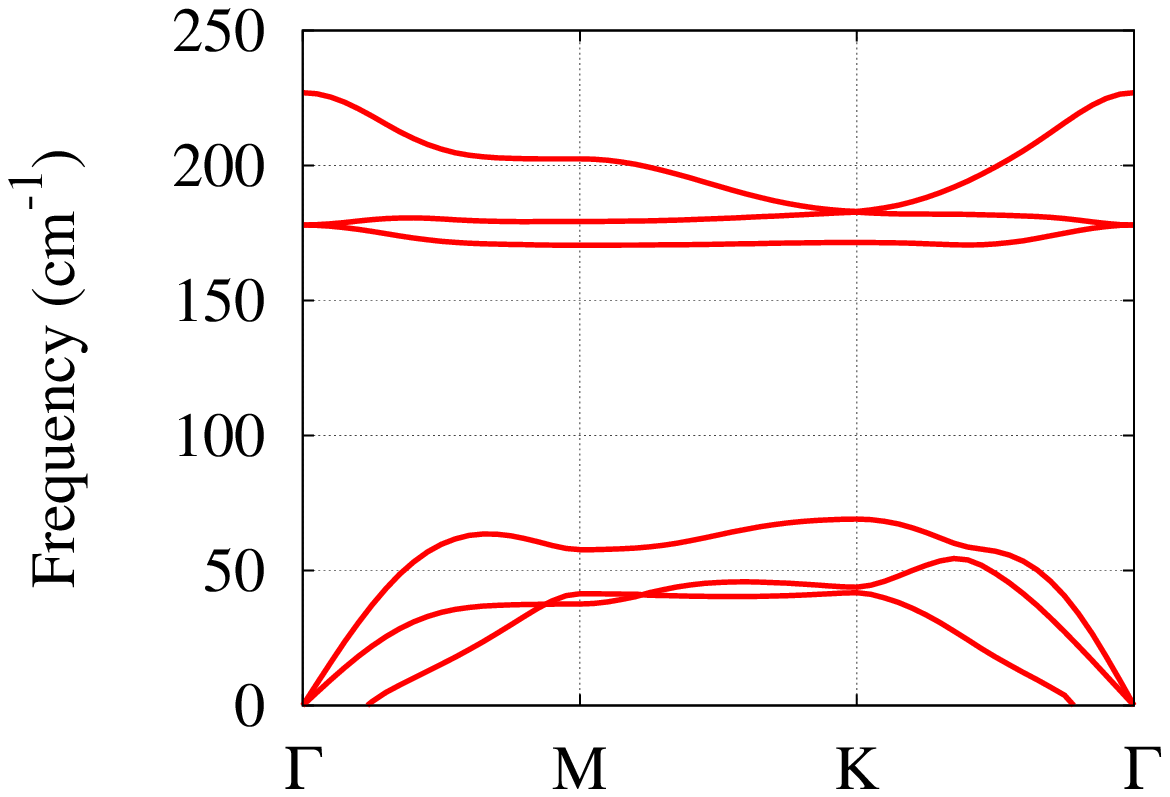}}
\subfigure[ {  } SbBi ]{\label{phonon_SbBi}
\includegraphics[width=5cm]{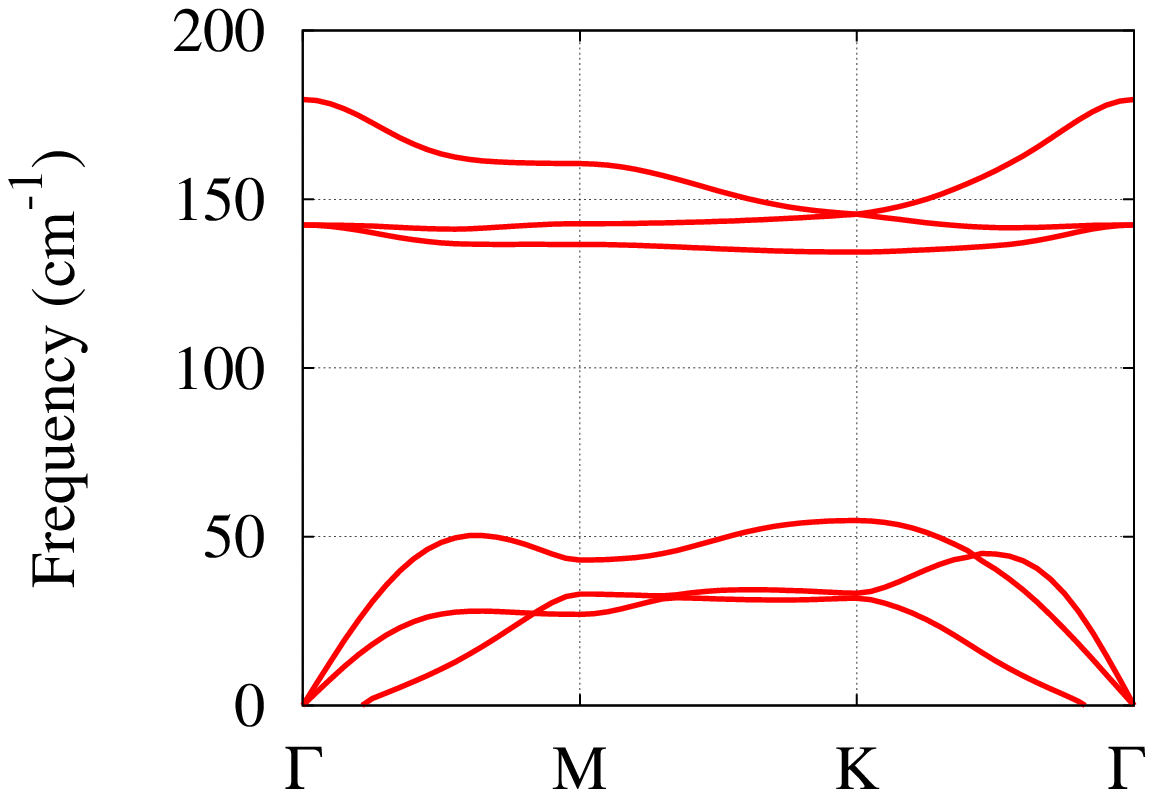}}
\caption{\label{phono} Phonon dispersion curves of buckled honeycomb structures of group-V binary compounds. }
\end{figure}

To check the stability, we calculate the phonon dispersion of these monolayer binary compounds. Absence of negative frequencies of the calculated phonon dispersion indicates the stability of monolayer PAs, PSb, PBi, and AsSb (see Figure \ref{phono}). Note that although ZA branch (out of plane acoustical modes) becomes soft and get imaginary frequencies near $\Gamma$ point for AsBi and SbBi, it is believed such instability can be removed by the defects\cite{cahangirov_two-_2009,sahin_monolayer_2009}. On the other hand, the ZA imaginary frequencies around  $\Gamma$ point also depends on the mesh size used in the calculations. It may be an artifact of the mesh size\cite{cahangirov_two-_2009,sahin_monolayer_2009}. The MD simulations also show the stability of these 2D compounds because no structural collapse happens during the simulations (see Supporting Information for more detail). 

\begin{figure}[htbp]
\subfigure[{  } PAs  ]{\label{band_PAs}
\includegraphics[width=5cm]{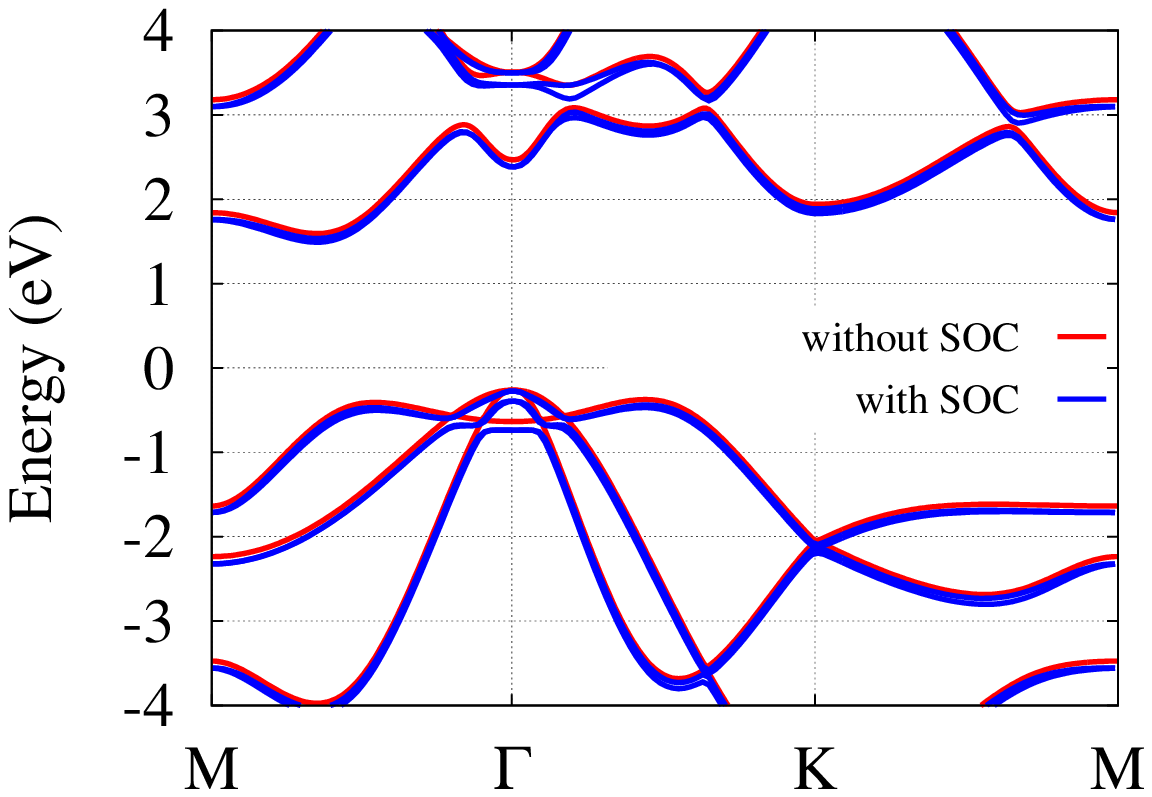}}
\subfigure[ {  } PSb ]{\label{band_PSb}
\includegraphics[width=5cm]{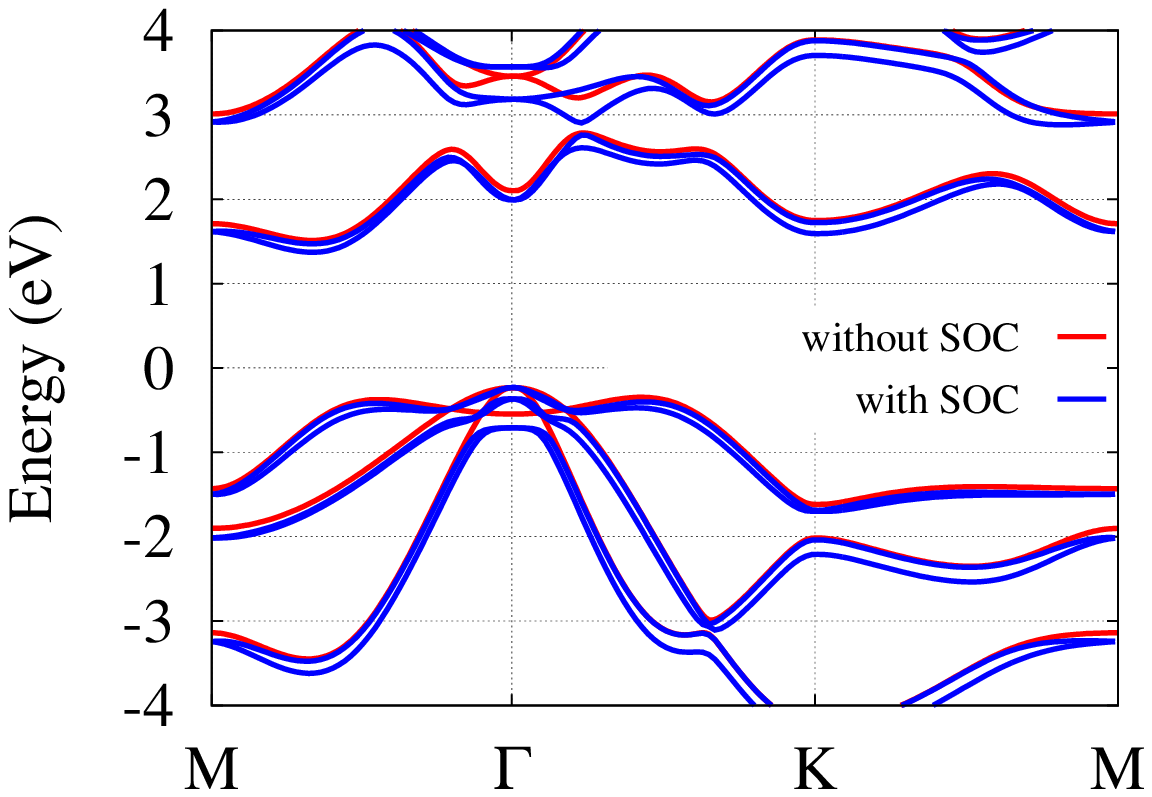}}
\subfigure[ {  } PBi ]{\label{band_PBi}
\includegraphics[width=5cm]{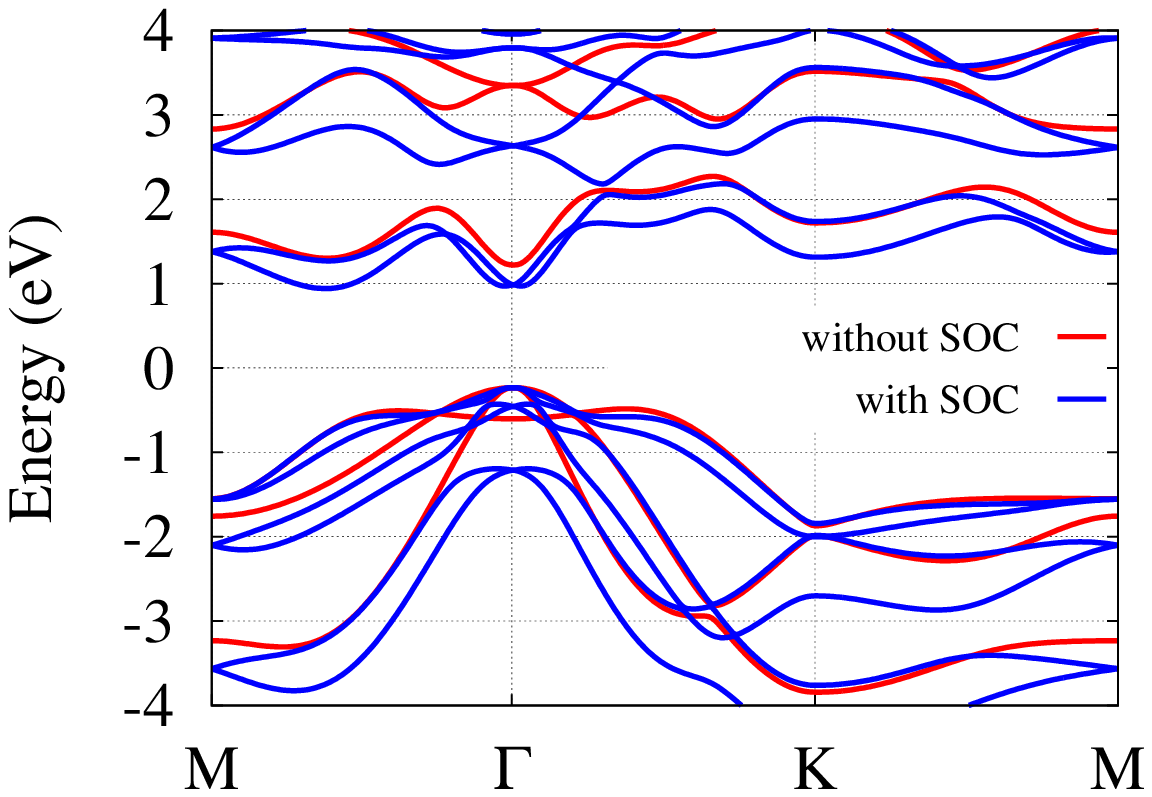}}
\subfigure[{  } AsSb ]{\label{band_AsSb}
\includegraphics[width=5cm]{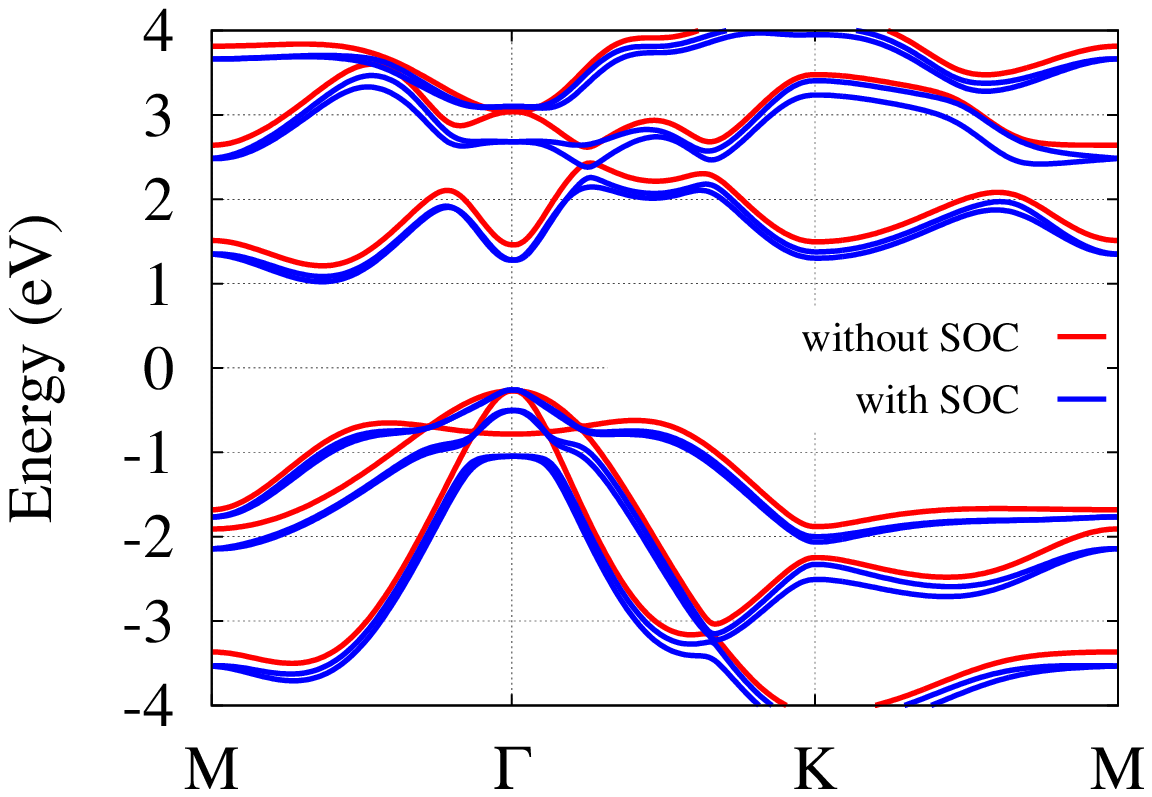}}
\subfigure[ {  } AsBi ]{\label{band_AsBi}
\includegraphics[width=5cm]{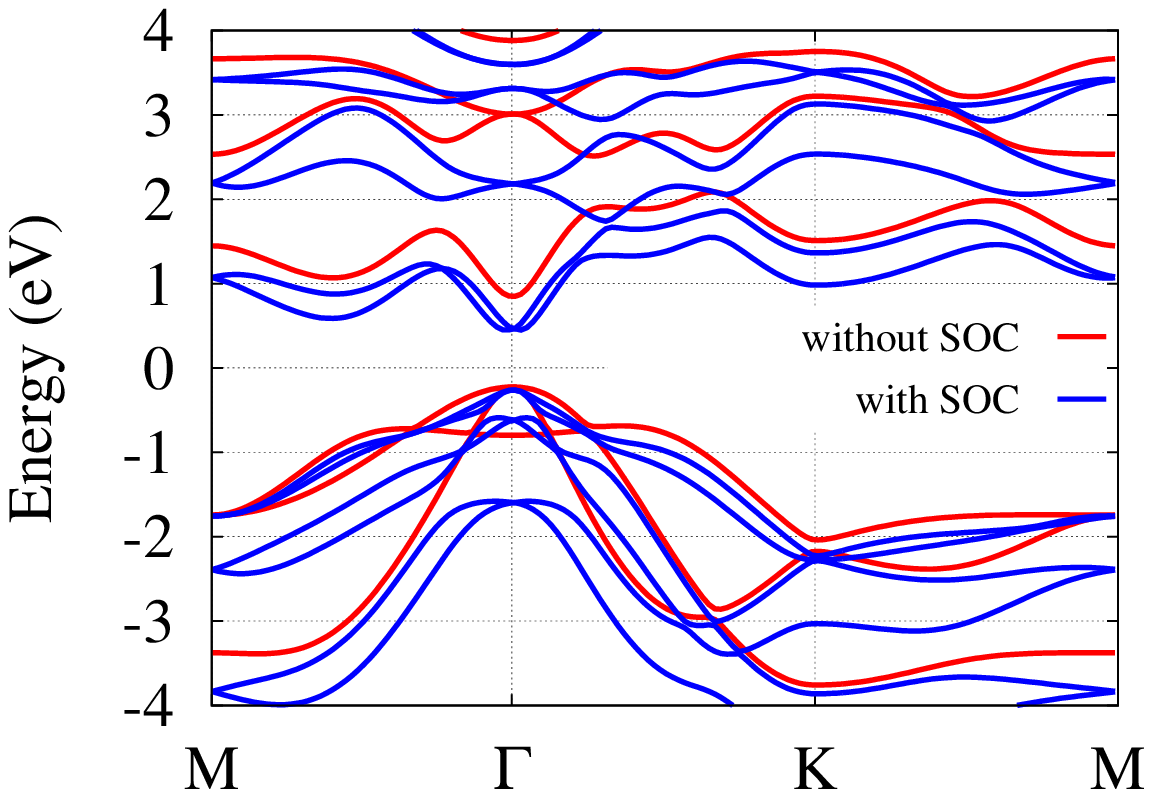}}
\subfigure[{  } SbBi ]{\label{band_SbBi}
\includegraphics[width=5cm]{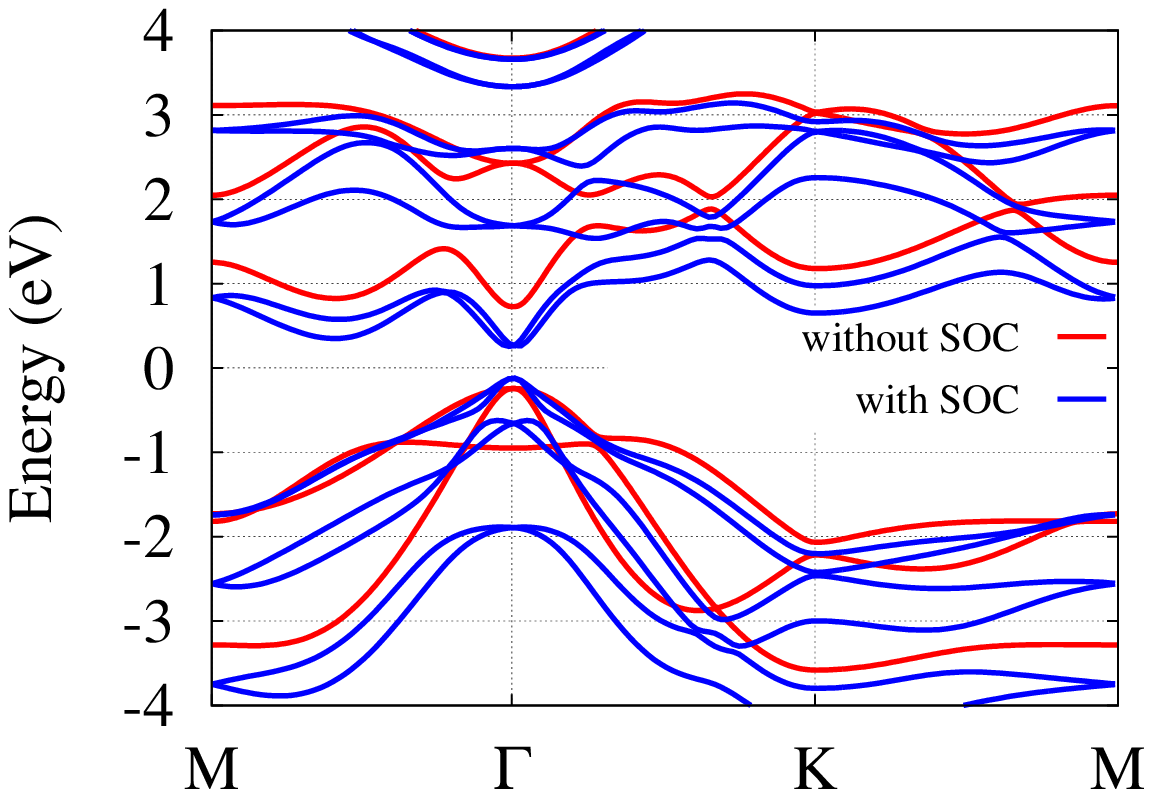}}
\caption{\label{band_AB} Energy bands of binary compounds with buckled honeycomb structure.}
\end{figure}

Figure \ref{band_AB} shows band structures of these binary compounds. The band structures of monolayer phosphorous, arsenic, antimony, and bismuth are also given for comparison (see Supporting Information for more detail).
The calculated band structures (without SOC) show similarities between different systems. This is not surprising, since these elements are chemically similar. Taking into account SOC, the number of bands of binary compounds is doubled due to the nonsymmetric structure and SOC. Note that SOC just modifies the band gaps but does not change band orders for these binary compounds, P, As, and Sb. However, the band structure of Bi differs from others when SOC is included. The parity of the states that form the valence and conduction bands is reversed at $\Gamma$ point. This exchange of bands is the consequence of the increased SOC in Bi, and Bi is reported a 2D topological insulator\citep{liu_stable_2011}. On the contrary, other 2D group-V elemental materials and binary compounds with the same buckled honeycomb structure are conventional insulators. Take monolayer AsSb for example, the calculated band structure (without SOC) shows it is a semiconductor with a indirect gap of 1.48 eV, while a gap of 1.73 eV at $\Gamma$. The valence band maximum (VBM) is at $\Gamma$ point, and the conduct band minimum (CBM) is at $k=(0.132, 0)$ point. The band structures (without SOC) of binary compounds containing Bi possess direct gaps, while others show indirect gaps. 

\begin{figure}[htbp]
\subfigure[]{\label{energy}
\includegraphics[width=5cm]{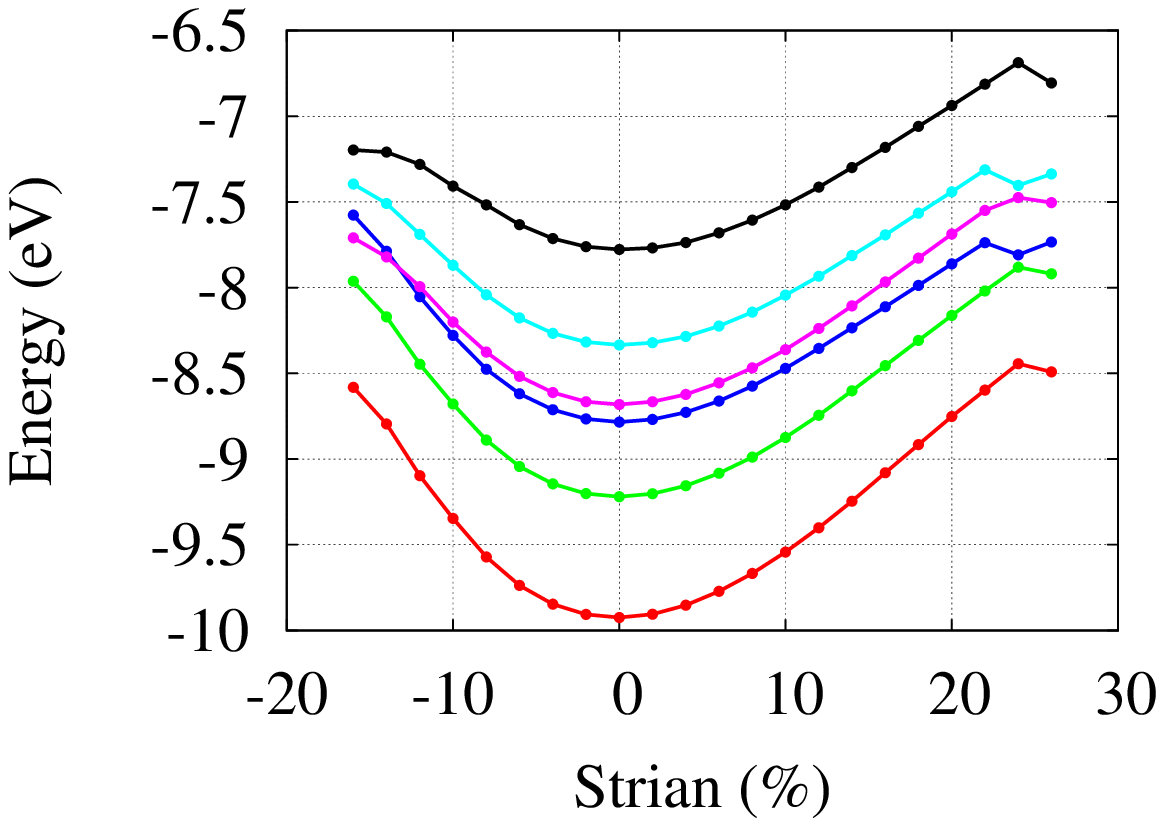}}
\subfigure[]{\label{stress}
\includegraphics[width=5cm]{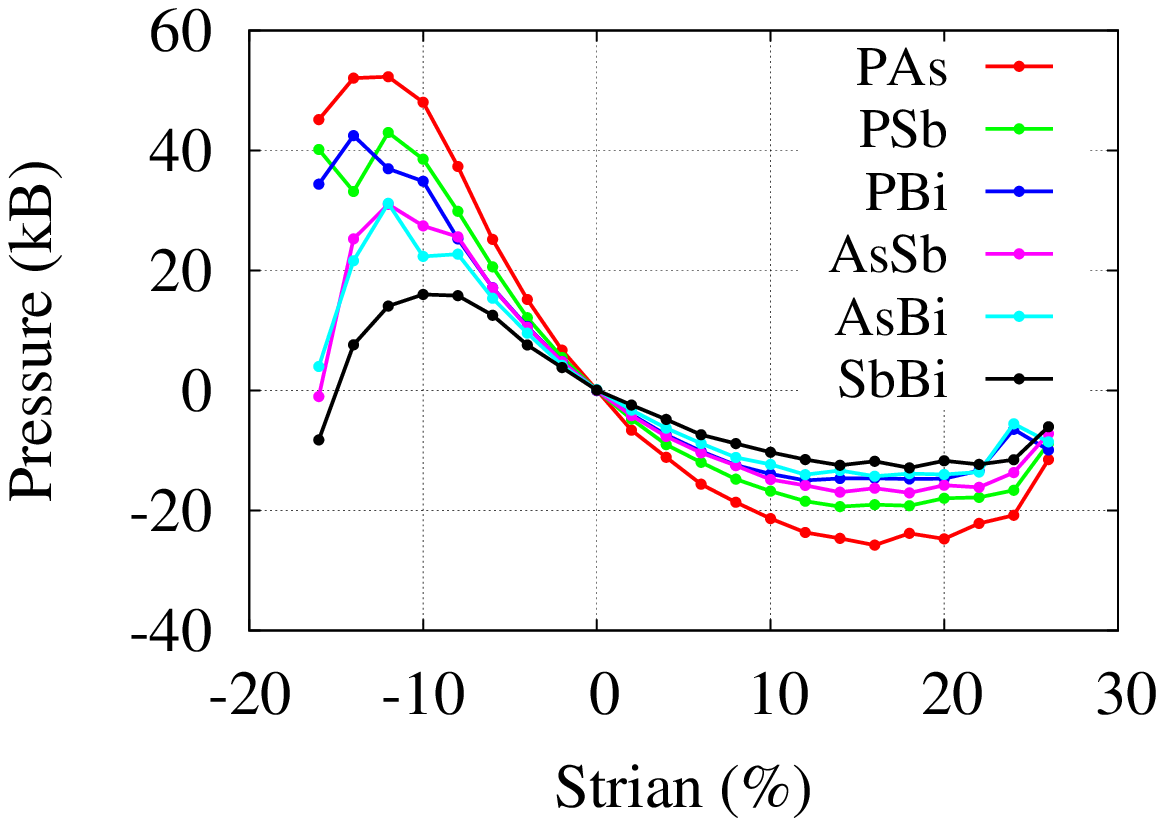}}
\subfigure[]{\label{parameters1}
\includegraphics[width=5cm]{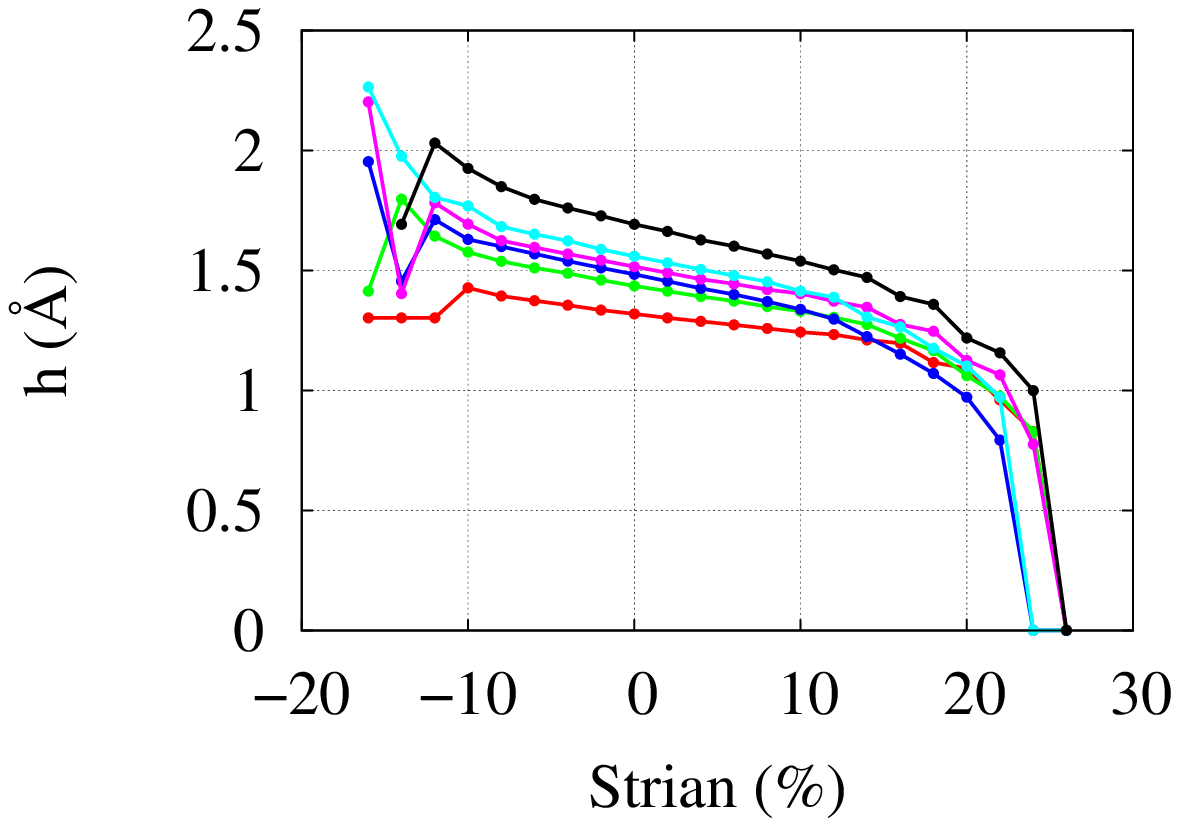}}
\subfigure[]{\label{parameters2}
\includegraphics[width=5cm]{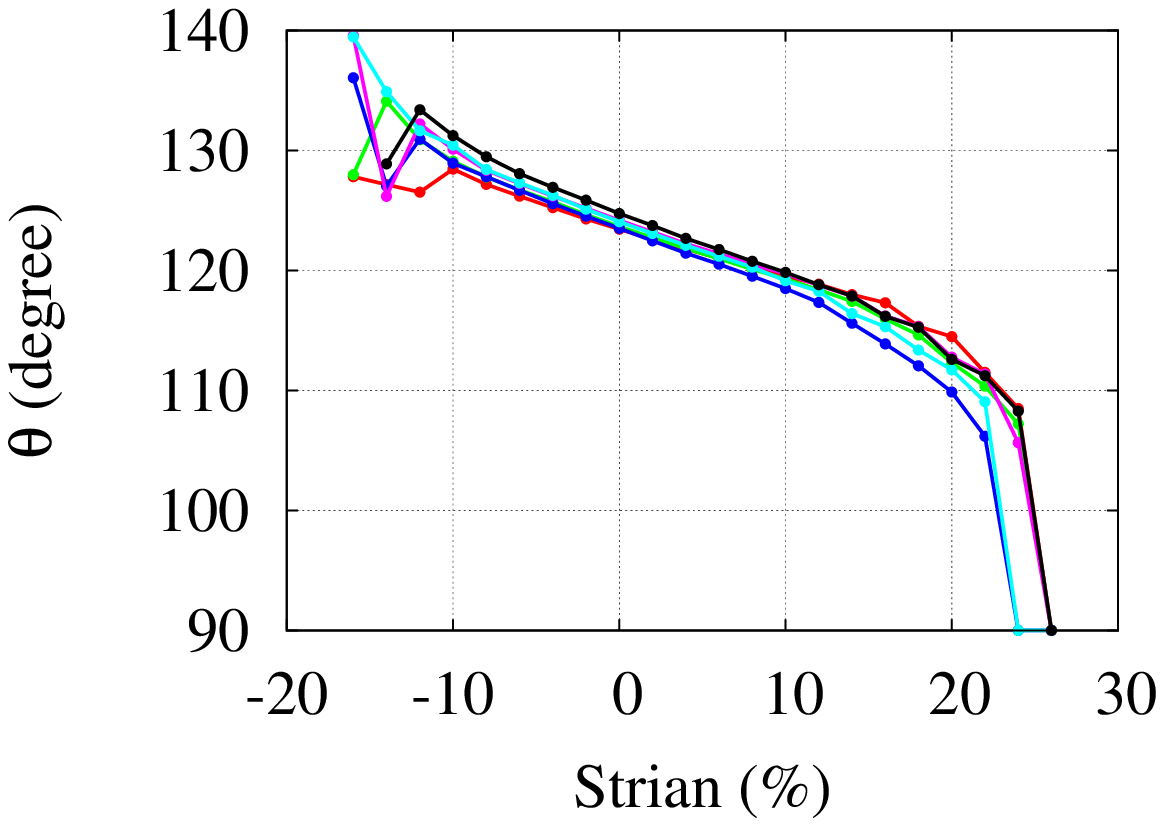}}
\subfigure[]{\label{gap}
\includegraphics[width=5cm]{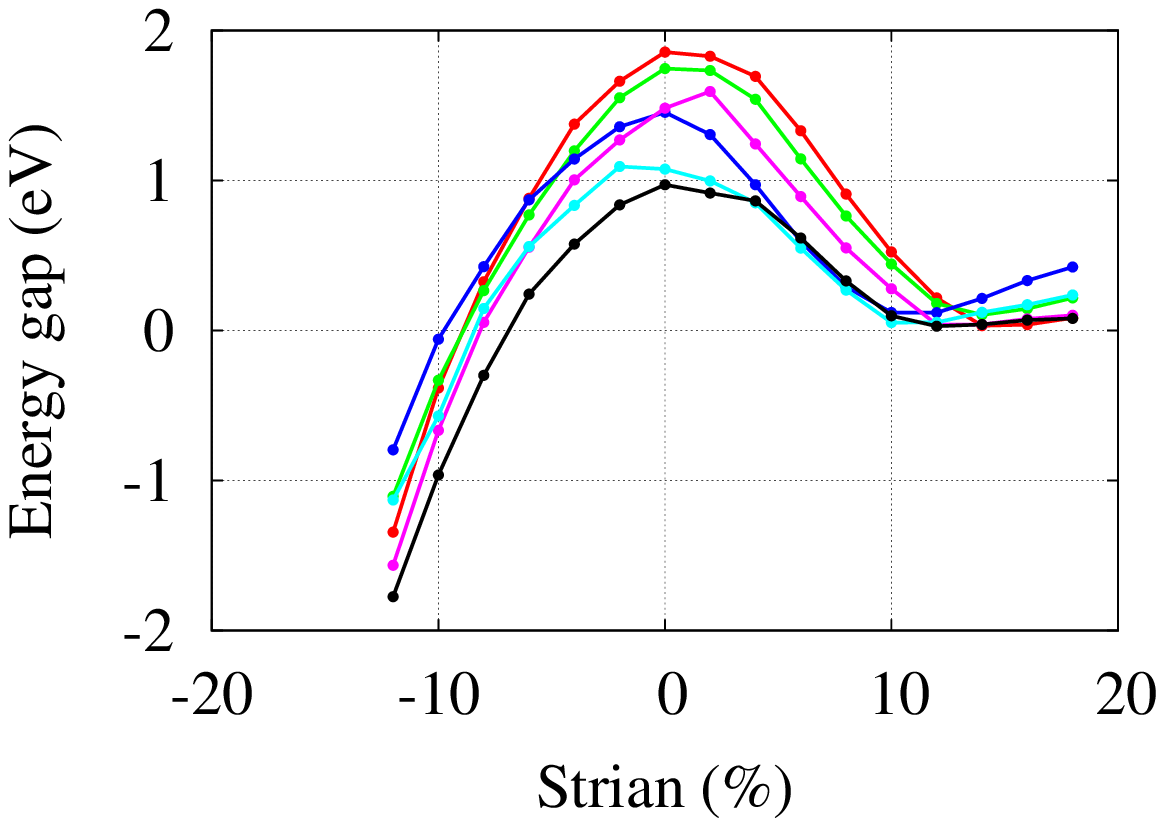}}
\subfigure[]{\label{gap_soc}
\includegraphics[width=5cm]{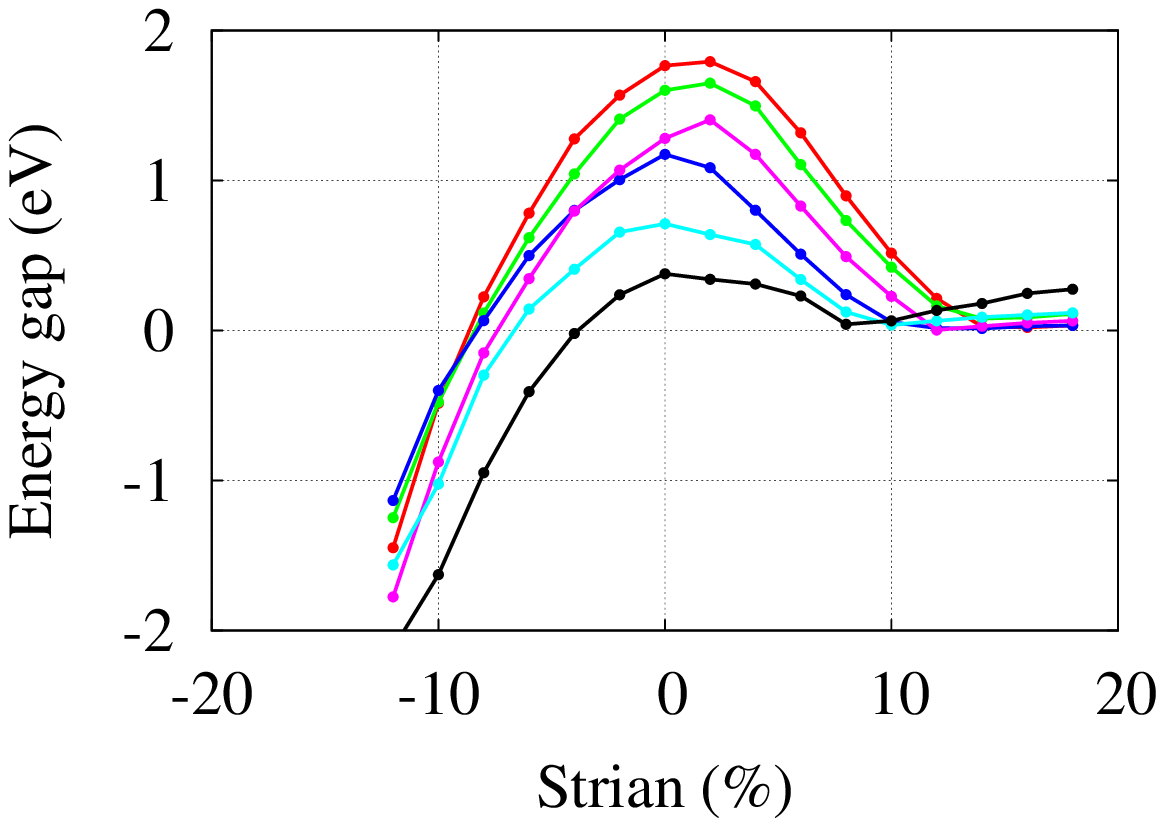}}
\caption{\label{parameters} Some parameters as functions of lattice strain. a: Total energy. b: In-plane pressure. c: Buckling parameter h. d: Angle $\theta$. e: Energy gap (without SOC). f: Energy gap (with SOC).}
\end{figure}

We perform strain engineering to these 2D binary compounds. Figure \ref{parameters} shows calculated energies, pressures, buckling parameters, and energy gaps as functions of the strain. From the place where the stress maximum(minimum) occur we can determine the stretchability of the material. Take AsSb for example, the stretchability is about $-12\%$ for press, and $18\%$ for stretch. During these domain, buckling parameters h and $\theta$ change almost linearly. When strain is larger than $26\%$, h drops to zero and $\theta$ equals $90^{\circ}$, and corresponding energy also changes rapidly. It means the system becomes planar honeycomb structure as graphene. Band gaps shown in Figure \ref{gap} (without SOC) and Figure \ref{gap_soc} (with SOC) indicate that the system is metallic when strain is between $-12\%$ and $-8\%$; when strain is between  $-8\%$ and $2\%$, the system is a indirect insulator; the system becomes a direct semiconductor with a gap at $\Gamma$ when strain is larger than $2\%$. The band gap decreases with increasing strain and closes at about $12\%$ strain, then reopen when strain is larger than $12\%$. Band structures of other binary compounds as well as P, As, and Sb under strain have similar characteristics (see Supporting Information for more detail). 

To get a better understanding of the mechanism underlying the band inversion, we investigate the band structure evolution at $\Gamma$ point. Take AsSb for example, As and Sb have five electrons in the outer shell. For each atom, two electrons occupy s orbital and three electrons occupy p orbitals. They form five energy bands below the Fermi energy. The two s bands are much lower than the three p bands. The three p bands of two atoms in the cell form six bands with three below the Fermi energy and three above the Fermi energy. According to our first-principles calculations, ignoring normalization factors, these six bands around $\Gamma$ point are mainly denoted by $|1\rangle=|p_{z}^{A}\rangle-|p_{z}^{B}\rangle$, $|2\rangle=|p_{z}^{A}\rangle+|p_{z}^{B}\rangle$ (in fact, there are some $|s^{A}\rangle$ and $|s^{B}\rangle$ components in $|2\rangle$), $|3\rangle=|p_{x}^{A}\rangle-|p_{x}^{B}\rangle$, $|4\rangle=|p_{y}^{A}\rangle-|p_{y}^{B}\rangle$, $|5\rangle=|p_{x}^{A}\rangle+|p_{x}^{B}\rangle$, and $|6\rangle=|p_{y}^{A}\rangle+|p_{y}^{B}\rangle$. A denotes As atom and B Sb atom. The $|1\rangle$, $|3\rangle$, and $|4\rangle$ are below the Fermi energy while $|2\rangle$, $|5\rangle$, and $|6\rangle$ are above the Fermi energy. When SOC is not included, $|3\rangle$ and $|4\rangle$ are degenerated, and so are $|5\rangle$ and $|6\rangle$. 

\begin{figure}[htbp]
\subfigure[{  }  ]{\label{band_p8}
\includegraphics[width=6cm]{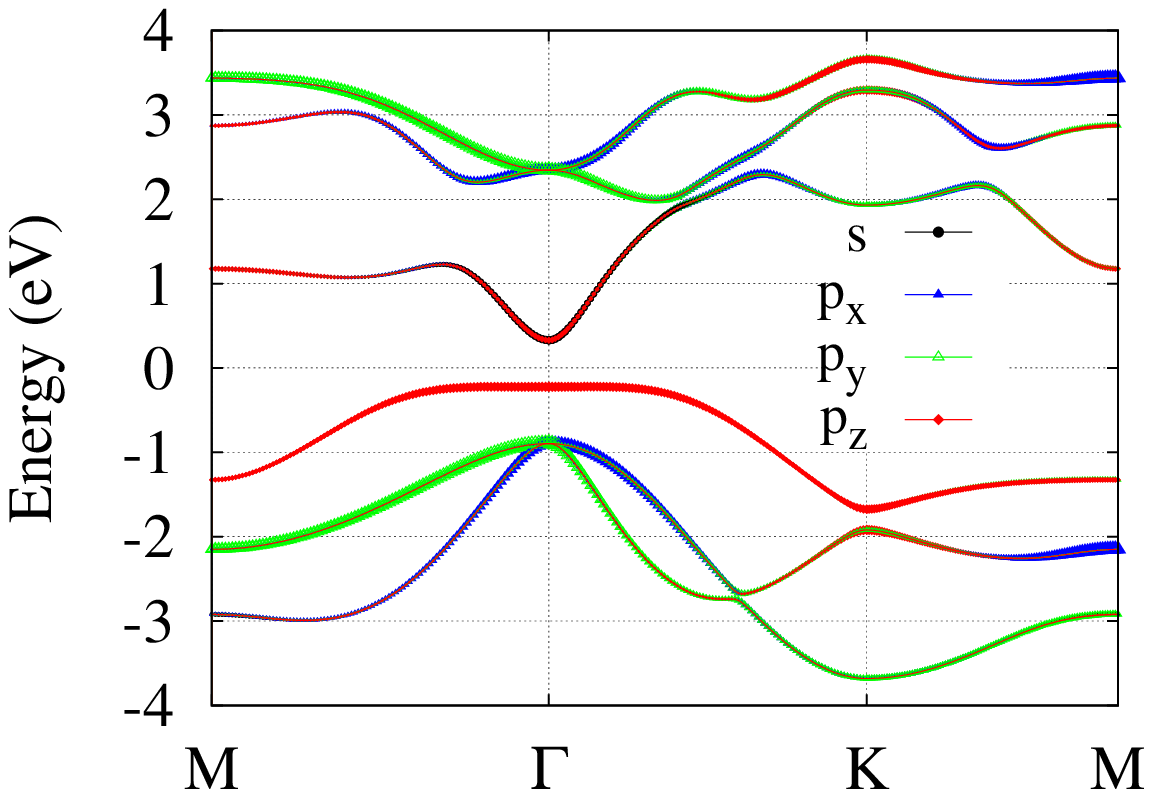}}
\subfigure[{  }  ]{\label{orbital_p8}
\includegraphics[width=1.5cm]{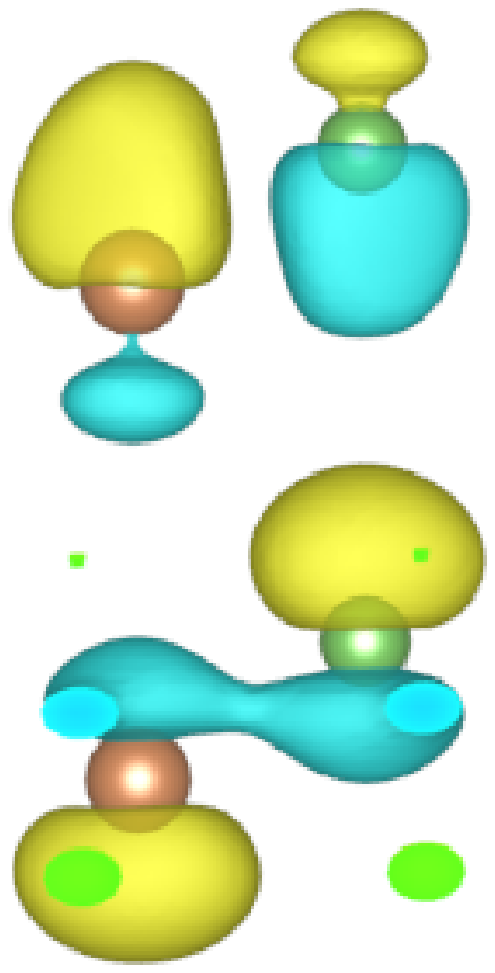}}
\subfigure[ {  }  ]{\label{band_p16}
\includegraphics[width=6cm]{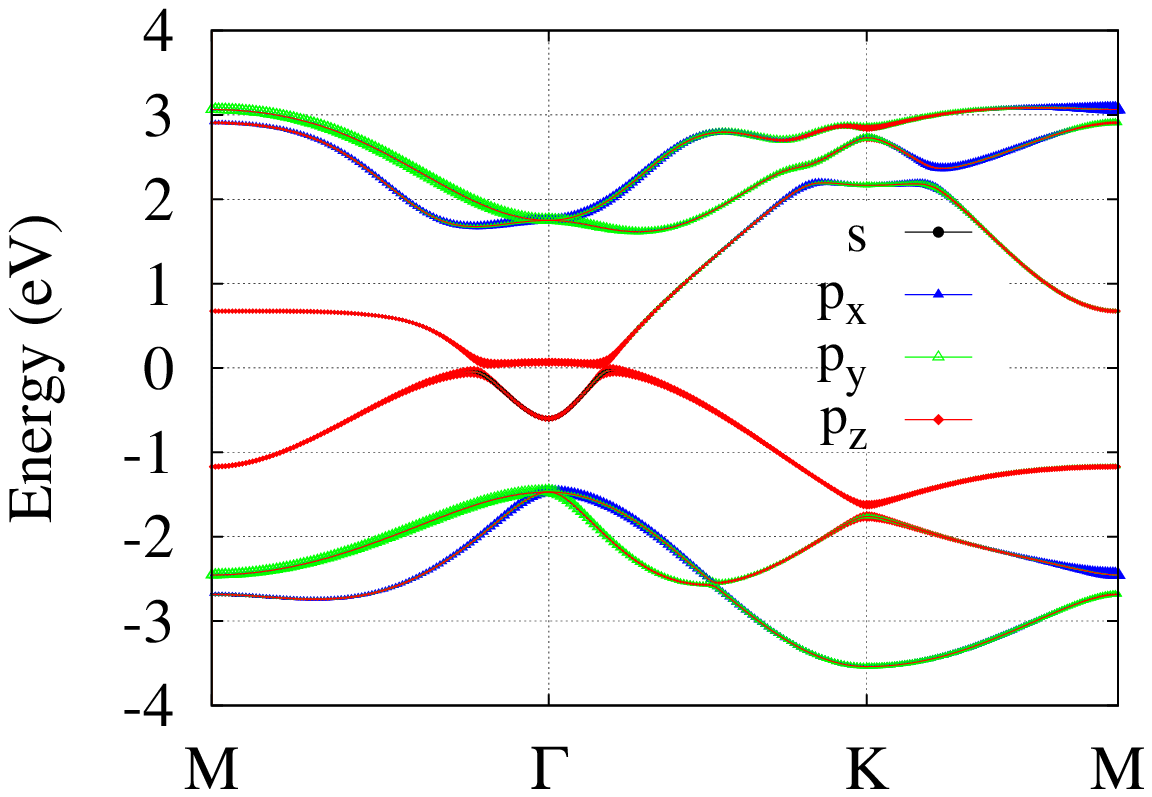}}
\subfigure[ {  }  ]{\label{orbital_p16}
\includegraphics[width=2cm]{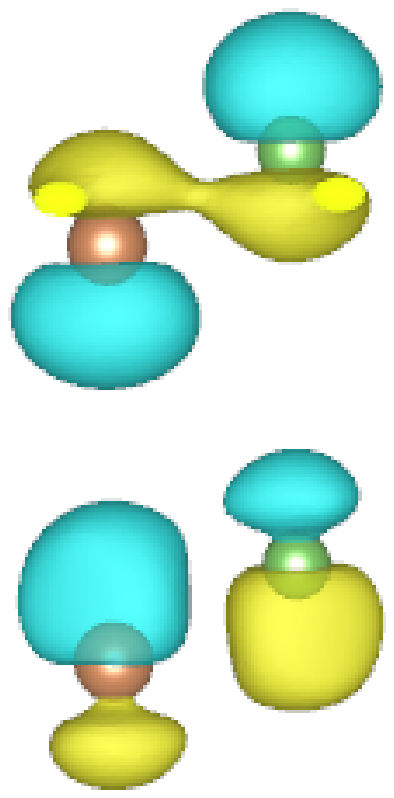}}
\caption{\label{orbital} Band structures of AsSb at different lattice strains, HOMO and LUMO orbitals at $\Gamma$ point. a: Band structure at $8\%$ lattice strain. b:  HOMO ($|1\rangle=|p_{z}^{A}\rangle-|p_{z}^{B}\rangle$) and LUMO ($|2\rangle=|p_{z}^{A}\rangle+|p_{z}^{B}\rangle-|s^{A}\rangle+|s^{B}\rangle$) orbitals at $\Gamma$ point at $8\%$ lattice strain. Green and brown spheres represent As atom and Sb atom, respectively. Yellow (green) color in orbitals means positive (negative) phase. c: Band structure at $16\%$ lattice strain. b:  HOMO ($|2\rangle$) and LUMO ($|1\rangle$) orbitals at $\Gamma$ point at $16\%$ lattice strain. }
\end{figure}

Under ambient conditions, $|1\rangle$ is below $|3\rangle$ and $|4\rangle$. When strain reaches up to $2\%$, the indirect gap becomes a direct gap. When strain is $3\%$, $|1\rangle$, $|3\rangle$, and $|4\rangle$ have the same energy. When strain is beyond $3\%$, the two bands below and above the Fermi energy at $\Gamma$ point is constructed by $|1\rangle$ and $|2\rangle$, respectively. Figure \ref{orbital} shows the band structures of monolayer AsSb at strain of $8\%$ and $16\%$ respectively, and the band composition. We see the band inversion at $\Gamma$ according to Figure \ref{band_p8} and Figure \ref{band_p16}. Because h is relatively large when strain is small, the low half of $|p_{z}^{A}\rangle$ is close to the upper half of $|p_{z}^{B}\rangle$, their coupling is bonding when they are out of phase, while their coupling is anti-bonding when they are in phase. Therefore $|1\rangle$ has a lower energy than $|2\rangle$. Figure \ref{orbital_p8} shows $|1\rangle$ and $|2\rangle$, corresponding to VBM and CBM, respectively. In other words, $|1\rangle$ is the highest occupied molecular orbital (HOMO) at $\Gamma$ point, while $|2\rangle$ is the lowest unoccupied molecular orbital (LUMO) at $\Gamma$ point. Note that $|2\rangle$ shows $|s^{A}\rangle$ and $|s^{B}\rangle$ components. With further increase of strain, h decreases and the low half of $|p_{z}^{A}\rangle$ is move from upper half of $|p_{z}^{B}\rangle$ to low half of $|p_{z}^{B}\rangle$, then energy of $|1\rangle$ increases and $|2\rangle$ decrease. When strain is about $12\%$, $|1\rangle$ and $|2\rangle$ have the same energy and the gap closes. When strain is larger than $12\%$, the low half of $|p_{z}^{A}\rangle$ is closer to the low half of $|p_{z}^{B}\rangle$ because of small h, then their coupling is bonding when they are in phase, while their coupling is anti-bonding when they are out of phase, therefore $|1\rangle$ has a higher energy than $|2\rangle$. Now $|2\rangle$ is HOMO at $\Gamma$ point, while $|1\rangle$ is LUMO at $\Gamma$ point, and the band inverts, as shown in Figure \ref{band_p16} and Figure \ref{orbital_p16}. 

When A and B are the same element such as P, As, and Sb monolayer, the system has inversion symmetry. $|1\rangle$ and $|2\rangle$ orbitals have even and odd parity, respectively. The $\mathbb{Z}_{2}$ invariant can be determined from the parities of the occupied states at the time-reversal invariant momenta in the Brillouin zone\cite{fu_topological_2007}. There are four time-reversal invariant momenta for buckled honeycomb structure, i.e., one $\Gamma$ and three M points, as shown in Figure \ref{structure:c}. The band inverts at $\Gamma$ means topological invariants $\mathbb{Z}_{2}$ changes. Directly parity counting at time reversal invariant momenta shows that the $\mathbb{Z}_{2}$ changes from 0 to 1 for monolayer P, As, and Sb, when stain is larger than $12\%$ (see Supporting Information for more detail). We note that As and Sb are reported to exhibit topological phase transition at tensile strain $6\%$\cite{huang_strain_2014}. However, this critical value is estimated according to bulk As and Sb lattice constants $a$, which is about $6\%$ longer than that of corresponding 2D lattice constants\cite{chuang_tunable_2013}. Taking monolayer As and Sb lattice constants as bases, the critical value should be $12\%$.    

When A and B are different elements, the inversion symmetry is broken. Then  $|1\rangle$, $|2\rangle$, and other states, are not the eigenstates of parity. Then it is not possible to determine the $\mathbb{Z}_{2}$ invariant from the symmetry of the occupied states at time-reversal invariant momenta. To check the topological invariants, we adopt the method proposed by Soluyanov and Vanderbilt\cite{soluyanov_computing_2011}. The method is to track the largest gap between Wannier charge centers (WCCs). The path following the largest gap between WCCs cross the WCCs bands a number of times that is equal, mod 2, to the $\mathbb{Z}_{2}$ invariant. 
The calculated $\mathbb{Z}_{2}$ indices show AsSb is a conventional insulator under $8\%$ strain and a topological insulator under $16\%$ (see Supporting Information for more detail). 

The mechanism of the band inversion at $\Gamma$ point is the different response of energies of $|1\rangle$ and $|2\rangle$ to lattice strain. The origin of the nontrivial topology in buckled honeycomb structures of group-V binary compounds (as well as monolayer P, As, and Sb with the same structure) results from the band inversion  due to lattice strain.  It is neither like the band inversion at $\Gamma$ point in monolayer Bi, HgTe quantum well, and $Bi_{2}Se_{3}$, which is due to SOC, nor like that in silicene and germanene, which results from massive Dirac cone and there is no band inversion. In the case of these 2D honeycomb structures of group-V binary compounds, SOC just modifies energy gaps, as shown in Figure \ref{gap} and Figure  \ref{gap_soc}. For example, monolayer AsSb, PBi, and SbBi exhibit topological energy gaps of 77, 333, and 69 meV under $16\%$ lattice strain without including SOC, respectively. Taking into account SOC, the corresponding energy gaps are 50, 29, and 247 meV, respectively.     

Note that the existence of $|s^{A}\rangle$ and $|s^{B}\rangle$ components in $|2\rangle$ does not change the mechanism of the band inversion mentioned above. However, it makes the difference between band dispersions of $|1\rangle$ and $|2\rangle$ near $\Gamma$ point. A remarkable feature can be seen that the $|1\rangle$ band shows more flatter with increasing strain, in particular when it gets close to the Fermi energy. In fact, the $|1\rangle$ band around $\Gamma$ point can be approximately given by (see Supporting Information for more detail)
\begin{equation}
\label{h11_pectrum}
\epsilon(k)=-3[(V_{pp\sigma}-V_{pp\pi})\cos^{2}\theta+V_{pp\pi}](M-Bk^{2})=E_{1}(M-Bk^{2})
\end{equation}
When the energy of $|1\rangle$ equals to that of $|2\rangle$, we have $E_{1}=0$. It means the valency band becomes flat. On the contrary, the existence of $|s^{A}\rangle$ and $|s^{B}\rangle$ components in $|2\rangle$ makes its band dispersive. In recent years, lattice models with flat bands have attracted attention for a number of reasons, among them are enhanced interaction effects\cite{hu_topological_2011}.

\begin{figure}[htbp]
\subfigure[{  } ]{\label{ribbon_p8}
\includegraphics[width=5cm]{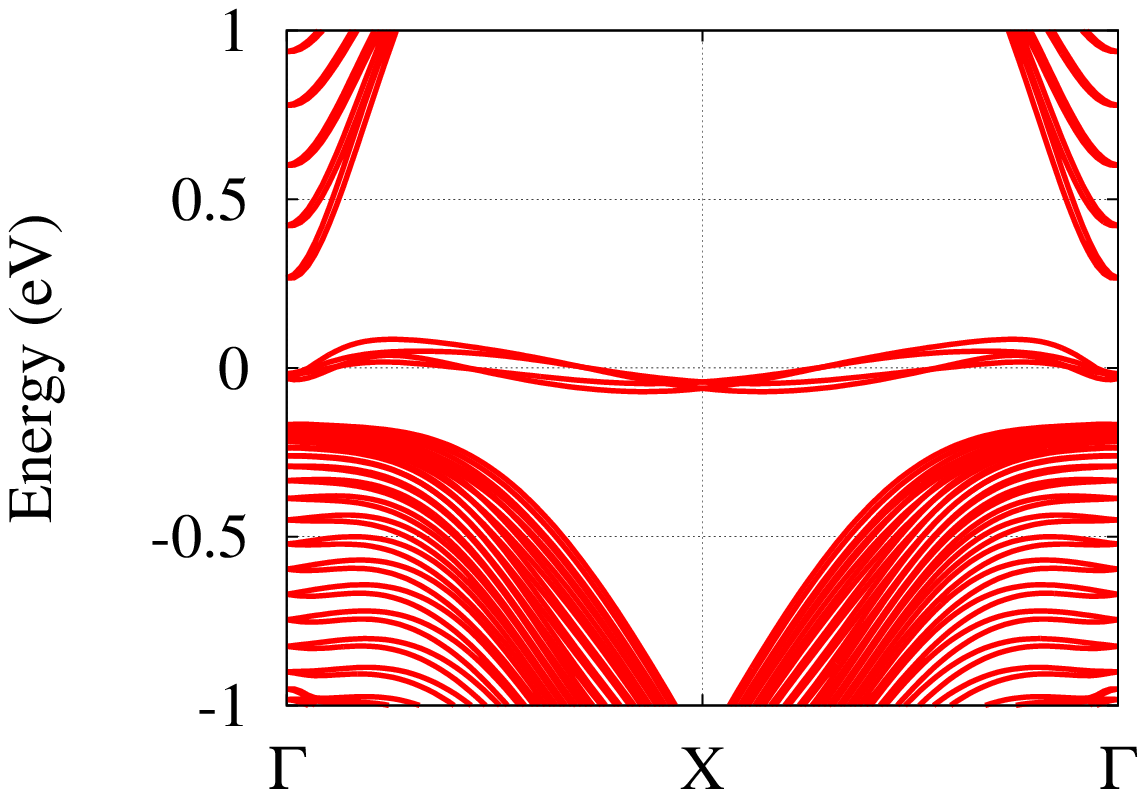}}%
\subfigure[{  }  ]{\label{ribbon_p161}
\includegraphics[width=5cm]{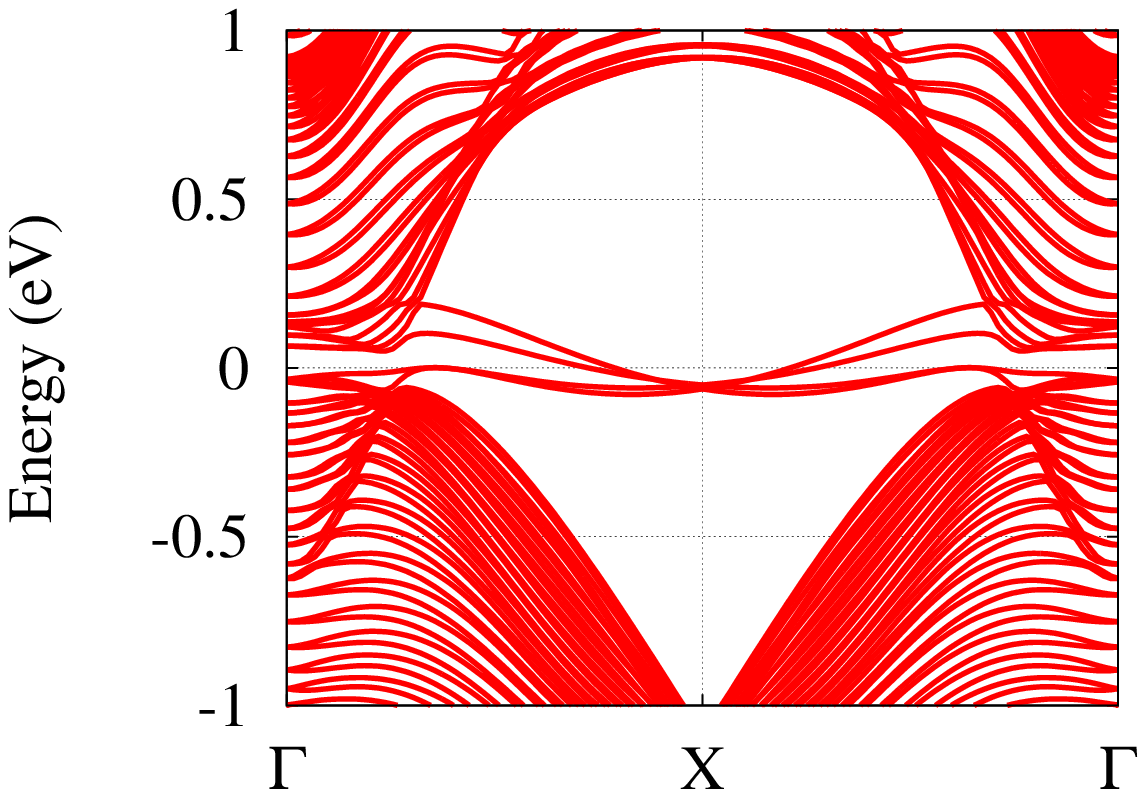}}
\subfigure[{  }  ]{\label{ribbon_p161}
\includegraphics[width=5cm]{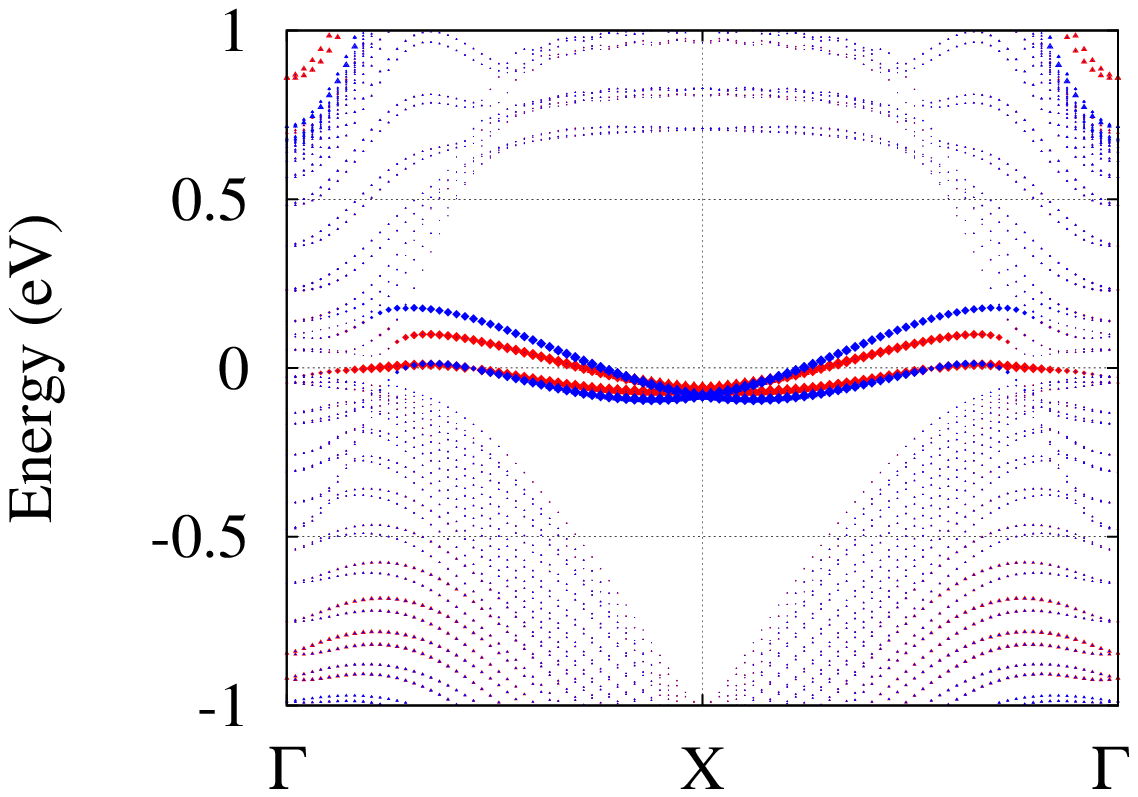}}
\caption{\label{ribbon} Band structures of AsSb nanoribbon under different lattice strains. a: Band structure at $8\%$ lattice strain. b: Band structure at $16\%$ lattice strain. c: The same band structure as shown in b. Red edge state comes from As atoms at one edge of the ribbon, while blue edge state comes from Sb atoms at the other edge of the ribbon.}
\end{figure}

The existence of topological protected gapless edge states is one of the most important consequences of 2D topological insulators. To verify the edge states, we construct zigzag ribbons with 48 atoms in a unit cell (about 84.40 \AA{} and 90.65 \AA{} wide), under strain $8\%$ and $16\%$, respectively.
Figure \ref{ribbon} shows the calculated band structures. We observe both ribbons show band gaps, and inside the gap, as expected, we observe four states, corresponding to two spin-splitting edge states from either side of the ribbon inside bulk gap.
AsSb ribbon has topologically protected gapless edge states under $16\%$. Although gapless edge states are also shown under $8\%$, they are not topologically protected. 

In conclusion, we predict that monolayer honeycomb structures of group-V binary compounds are stable according to phonon dispersion and molecular dynamical simulations. These new 2D materials are semiconductors under ambient conditions. Monolayer PBi, AsBi, and SbBi  possess direct gaps, while PAs, PSB, and AsSb possess indirect gaps. Under lattice strain, they show topological phase transitions according to calculated $\mathbb{Z}_{2}$ invariants.  By constructing ribbons, we show topologically protected edge states exist at nontrivial topology phase. Analysis of the band structure evolution at $\Gamma$ point indicates that the band inversion is due to lattice strain and irrelevant to SOC. This mechanism of topological transition under lattice strain is different from that of previously studied 2D TIs, in which SOC is indispensable for nontrivial band topology. Therefore it is possible to seek nontrivial band topology in materials composed of constituents with lighter atoms and thus smaller SOC. We hope that our work will promote the research aiming at the synthesis of these new 2D materials and the search for new 2D TIs. 

\begin{acknowledgement}
We wish to acknowledge the support of the National Natural Science
Foundation No.11374373, Doctoral Fund of Ministry of
Education of China (No.20120162110020), and the Natural Science
Foundation of Hunan Province of China (No.13JJ2004).
\end{acknowledgement}


\clearpage

\end{document}